# Magnetic Particle Spectroscopy (MPS) with One-stage Lock-in Implementation for Magnetic Bioassays with Improved Sensitivities

Vinit Kumar Chugh[a,†], Kai Wu[a,†,*], Venkatramana D. Krishna[b], Arturo di Girolamo[a], Robert P. Bloom[a], Yongqiang Andrew Wang[c], Renata Saha[a], Shuang Liang[d], Maxim C-J Cheeran[b,*], and Jian-Ping Wang[a,*]

[a]Department of Electrical and Computer Engineering, University of Minnesota, Minneapolis, MN 55455, United States

[b]Department of Veterinary Population Medicine, University of Minnesota, St. Paul, MN 55108, United States

[c]Ocean Nano Tech LLC, San Diego, CA 92126, United States

[d]Department of Chemical Engineering and Material Science, University of Minnesota, Minneapolis, MN 55455, United States

**ABSTRACT:**
In recent years, magnetic particle spectroscopy (MPS) has become a highly sensitive and versatile sensing technique for quantitative bioassays. It relies on the dynamic magnetic responses of magnetic nanoparticles (MNPs) for the detection of target analytes in liquid phase. There are many research studies reporting the application of MPS for detecting a variety of analytes including viruses, toxins, and nucleic acids, etc. Herein, we report a modified version of MPS platform with the addition of a one-stage lock-in design to remove the feedthrough signals induced by external driving magnetic fields, thus capturing only MNP responses for improved system sensitivity. This one-stage lock-in MPS system is able to detect as low as 781 ng multi-core Nanomag50 iron oxide MNPs (micromod Partikeltechnologie GmbH) and 78 ng single-core SHB30 iron oxide MNPs (Ocean NanoTech). In addition, using a streptavidin-biotin binding system as a proof-of-concept, we show that these single-core SHB30 MNPs can be used for Brownian relaxation-based bioassays while the multi-core Nanomag50 cannot be used. The effects of MNP amount on the concentration dependent response profiles for detecting streptavidin was also investigated. Results show that by using lower concentration/amount of MNPs, concentration-response curves shift to lower concentration/amount of target analytes. This lower concentration-response indicates the possibility of improved bioassay sensitivities by using lower amounts of MNPs.

Magnetic particle spectroscopy (MPS) for magnetic bioassays was first reported in 2006.[1,2] It is a technology that derived from magnetic particle imaging (MPI), which relies on the nonlinear magnetization curves of magnetic nanoparticle (MNP) tracers for medical tomographic imaging.[3] While, on the other hand, MPS monitors the dynamic magnetic responses of MNPs in liquid phase and assists in the analysis of the nanoparticles' binding status. To be specific, MNPs dispersed in the liquid adds an additional degree of rotational freedom that allows for bioassays directly from liquid phase. Upon the application of external AC magnetic fields (also called driving fields or excitation fields), the magnetizations of MNPs follow the field direction through a Brownian relaxation process, which is a physical rotational motion of nanoparticles. The dynamic magnetic responses of MNPs can be transformed to real-time voltage signal and monitored by using a pair of pick-up coils. The signal spectrum contains higher harmonics that are uniquely generated by MNPs. MPS-based bioassays use these harmonics of oscillating MNPs as a measure of the rotational freedom, i.e., the bound status of MNPs to target analytes from liquid phase. With appropriate chemical modifications, MNPs can be surface functionalized with proteins (such as antibodies, antigens, streptavidin, biotin, etc.), nucleic acids (DNA and RNA), and polymers, customized according to different bioassay purposes.[4–6] These surface functionalized MNPs are nanoprobes that can bind to target analytes with high specificity and have shown great promise not only for MPS-based bioassays but also other magnetic bioassays and medical applications. Some target applications include magnetoresistive and magnetic impedance biosensors, and nuclear magnetic resonance biosensors.[7–17] In liquid phase MPS-based bioassays, the binding of MNPs to targe analytes will hinder or even block the Brownian relaxation of MNPs and thus, causes a phase lag between their magnetizations and the external AC magnetic fields. The binding of MNPs to analytes causes weaker dynamic magnetic responses of MNPs and, as a result, the harmonic amplitudes drop is expected. Thus, this assay mechanism allows for development of one-step, wash-free, and quantitative detection of target analytes directly in liquid phase.[18]

In the past decade, various MPS platform designs have been reported, such as two AC (also called frequency mixing) and one AC driving fields methods based on how the excitation fields are applied, as well as the surface- and liquid phase-based bioassays (based on how the MNP are bound).[11,12,18–25] However, a common issue with all MPS systems is the presence of feedthrough signal corresponding to the driving magnetic fields which can be orders of magnitude higher than that of the MNP signal and can be a limiting factor in the MPS-based bioassays. Modalities based on both active and passive cancellation for such signal have been explored.[26–30] In this present work, we are reporting a modified version of two AC magnetic fields-based MPS system with the addition of one-stage lock-in scheme for passive cancellation of feedthrough signal and improved detection sensitivity. The performance and sensitivity of this one-stage lock-in MPS system is firstly evaluated by determining the lowest amount of MNPs detectable in liquid phase. Then we explored the impact of MNP amounts on the ability of MPS system to detect varied concentrations of target analytes by using a streptavidin-biotin binding system in liquid phase.

- **MATERIALS AND METHODS**

**Materials.** The Nanomag50 MNPs are 50 nm superparamagnetic dextran iron oxide composite nanoparticles functionalized with biotin, with weight concentration of 5 mg/mL and particle concentration of 91.36 nM, purchased from micromod Partikeltechnologie GmbH (product no. 79-26-501). The SHB30 MNPs are 30 nm iron oxide nanoparticles functionalized with biotin, with weight concentration of 1 mg/mL and particle concentration of 34 nM, provided by Ocean NanoTech. Streptavidin from *Streptomyces avidinii* is purchased from Sigma-Aldrich (product no. S4762).

**Magnetic Property Characterization.** 10 μL of Nanomag50 and SH30 MNP suspensions are each transferred to filter paper and air-dried. Then the static magnetic hysteresis loops are measured by a Physical Properties Measurement System (PPMS, Quantum Design Inc.) to obtain the magnetic properties of these nanoparticles such as the saturation magnetization ($M_s$) and coercivity ($H_c$). The specific magnetic magnetizations (M, emu/g) of Nanomag50 and SHB30 MNPs under 500 Oe field are 18.2 emu/g and 29.2 emu/g, respectively. Nanomag50 MNPs show superparamagnetic properties with zero magnetic coercivity while, on the other hand, SHB30 MNPs show a coercivity field of 36 Oe. The hysteresis loops and magnetic properties of Nanomag50 and SHB30 MNPs are summarized in Supporting Information S1.

**MPS Measurements.** The dynamic magnetic responses of MNPs are characterized using a homebuilt one-stage lock-in MPS system. A fixed volume of MNP suspension (with or without protein analytes) is added to a glass vial that can fit into the pick-up coils. AC magnetic fields are applied by drive coils and the dynamic magnetic responses are sensed by pick-up coils in the form of real-time voltage signal. Several MPS readings are taken from each vial and each reading consists of 170k discrete-time voltage samples from which the higher harmonics are extracted.

- **RESULTS AND DISCUSSIONS**

**Magnetic Particle Spectroscopy (MPS) Platform.** The MPS platform consists of two main parts, namely, magnetic field generation and MNP signal decoding. The MPS system operates on frequency mixing modality where a magnetic field consisting of one high frequency ($f_H$) and one low frequency ($f_L$) components is applied and the dynamic magnetic response of MNPs is recorded in accordance with the principle of Faraday's law. The experiments in this present work are conducted with the dual-frequency magnetic field having sinusoidal components of 5 KHz, 25 Oe for the high frequency driving field and 50 Hz, 250 Oe for the low frequency driving field. A balanced set of pick-up coils is used as shown in Figure 1(a) for recording the magnetic response which in principle cancels out any EMF generated due to the applied magnetic field and permits exclusive recording of the magnetic response from MNPs. However, in practice, the feedthrough signal (i.e., EMF due to driving fields) tends to be a real problem. Figure 1(f) shows the FFT spectrum of MNP response and it can be clearly observed that the feedthrough signal corresponding to $f_H$ is two orders of magnitude higher than the 3rd harmonic responses of MNPs.

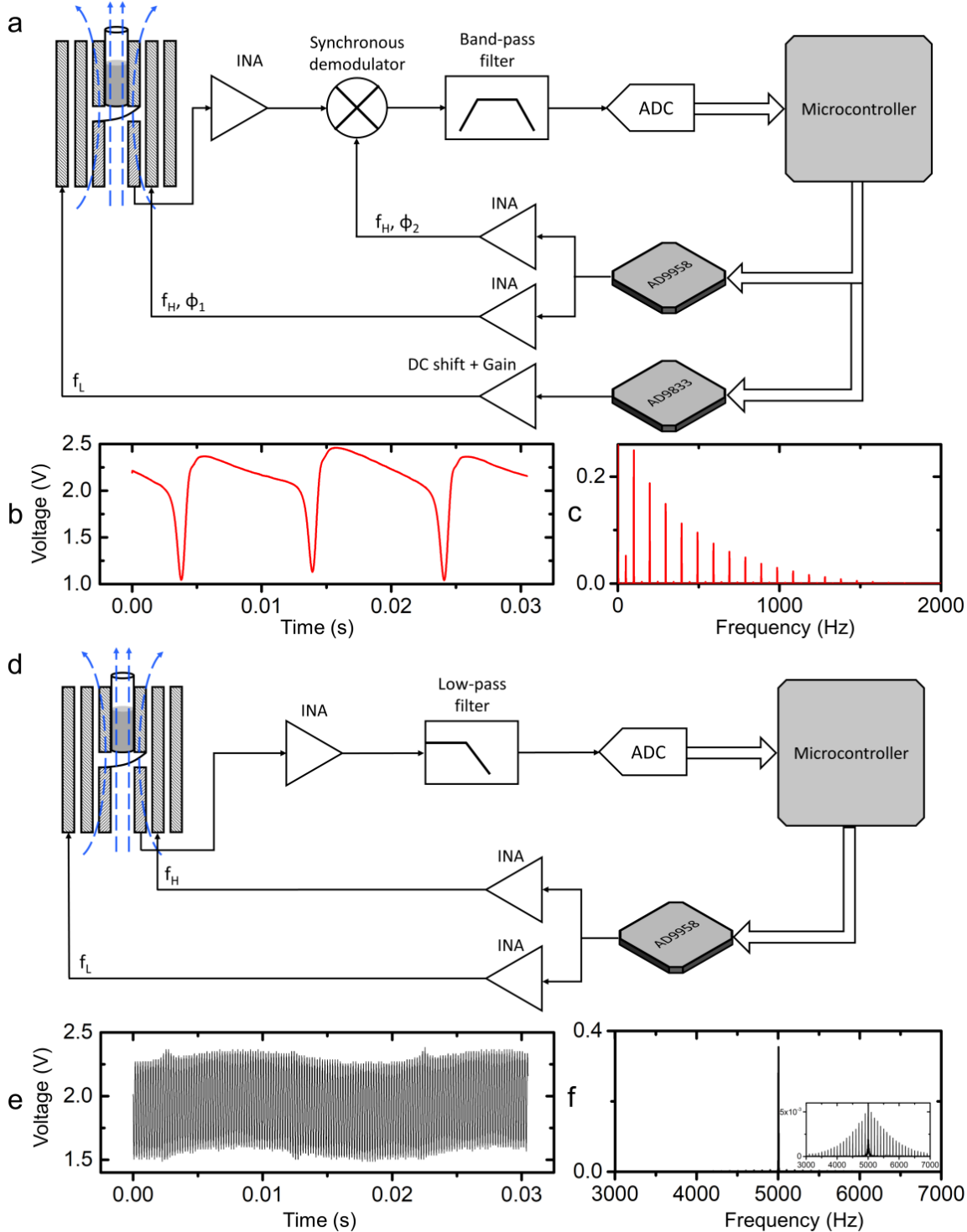

Figure 1. Schematic view of MPS signal flow with (a) and without the (d) one-stage lock-in implementation. Temporal-domain signal as sampled by ADC is shown in (b) and (e) whilst corresponding signal depiction in frequency-domain is shown in (c) and (f) for the lock-in and without lock-in based approaches.

Herein, a one-stage lock-in based approach is used to remove the feedthrough signals corresponding to driving field frequencies and to capture only the MNP responses. Figure 1(a) & (d) depicts schematic diagrams of the signal decoding topology with and without the one-stage lock-in implementation and the corresponding captured signals in temporal and frequency domains (Figure 1(b), (c), (e), and (f)). From Figure 1(c), we can clearly observe that the one-stage lock-in approach significantly removes the feedthrough signals. Another advantage that lock-in based approach provides is to reduce sampling frequency requirements. Experiments on a MPS system without lock-in design (Figure 1(d)) require a sampling frequency of up to 500 KSPS to obtain optimal SNR performance. However, the lock-in based approach shifts the MNP spectra to a lower frequency range and hence allows for better SNR performance with lower sampling rate (100 KSPS) that is easily adaptable to on a handheld system setting.

**Circuit Design of One-stage Lock-in MPS System.** The circuit for one-stage lock-in MPS system can be divided into 3 main parts: (1) power, (2) excitation coil driver, and (3) signal processing. Figure 2 depicts a simplified block diagram for the one-stage lock-in MPS system developed for this work.

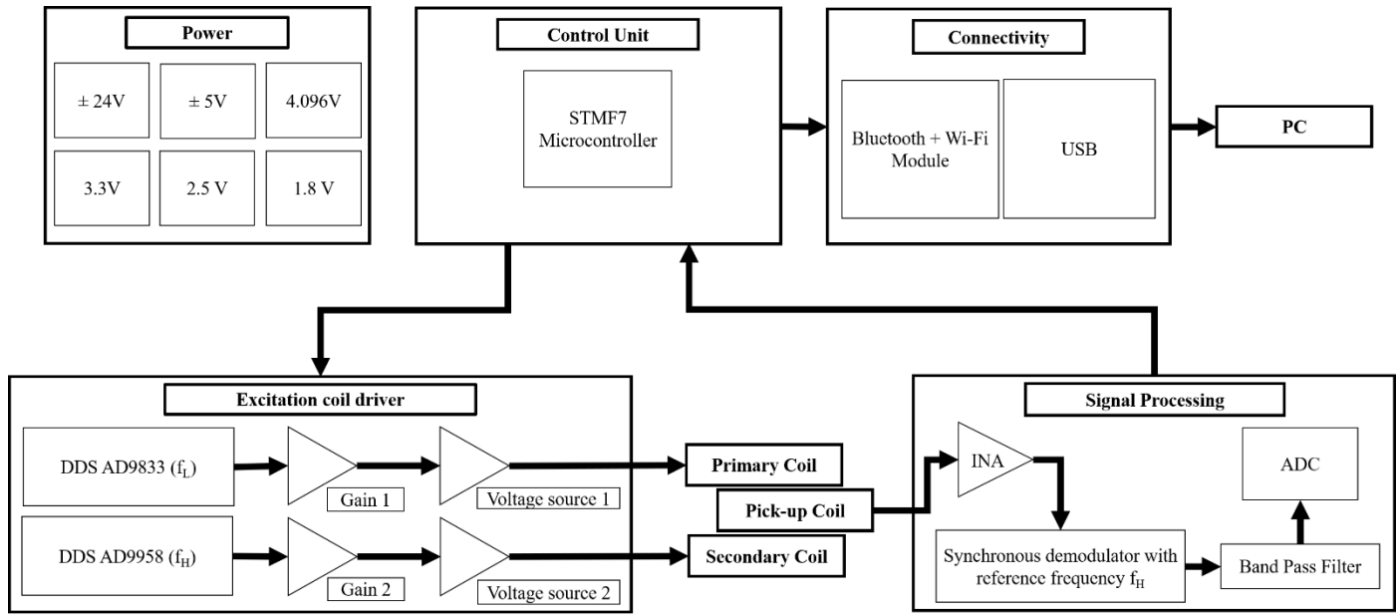

Figure 2. Block diagram of MPS system with one-stage lock-in implementation.

**Power Generation.** MPS device requires different line voltages for operation of different system blocks. Briefly, the system requires ±24 V for excitation coil voltage source driver, ±5 V for signal processing stages, 3.3 V for AD9833 and STMF7 microcontrollers, 4.096 V and 2.5 V for ADC reference and power correspondingly, and 1.8 V for the AD9958 DDS IC. The MPS device utilizes a 48 V wall adapter, GSM160A48-R7B (MEAN WELL Enterprises) as the main power source. The ±24 V supply is generated using a low-noise powerline split implementation using OPA549 (Texas Instruments), see Supporting Information S2 for details. The voltages 5 V, -5 V, 3.3 V, 2.5 V, and 1.8 V are generated using LDOs LT1117 (Linear Technology), LT1175 (Linear

Technology), TPS62177 (Texas Instruments), ADP1715-2.5 (Analog Devices), and ADP3338-1.8 (Analog Devices), respectively.

**Excitation Coil Driver.** The coil excitation circuit consists of sine wave generation followed by voltage source implementation. A 2-channel DDS IC AD9958 (Analog Devices) is used for generation of sinusoids for high-frequency ($f_H$) driving field and for phase shifted reference to the synchronous demodulator in signal processing stage. Differential output from AD9958 is passed through instrumentation amplifier INA128 (Texas Instruments) to convert into the single ended signal. DDS AD9833 (Analog Devices) is used for generation of sinusoid for low-frequency ($f_L$) driving field. An inverting amplifier topology utilizing OPA548 (Texas Instruments) is used for the voltage source implementation for driving the primary and secondary coils.

**Signal Processing.** The differential signal from pick-up coils is amplified using the precision instrumentation amplifier INA828 (Texas instruments) for removing the common mode noise. Signal at this stage consists of MNP response centered around 5 KHz excitation frequency as can be seen in Figure 1(f). The amplified signal is processed using a lock-in based implementation consisting of an AD630 synchronous demodulator (Analog Devices) with phase shifted 5 KHz ($f_H$) reference signal followed by a bandpass filter to reject signal images at 0 Hz and around 10 KHz. The band-pass filter is implemented using the Sallen-key scheme. The filtered signal is sampled using LTC2368-24, 24-bit SAR ADC (Linear Technology, Analog Device) at 100 KSPS sampling rate. STM32F747 microcontroller (STMicroelectronics) is used for handling and storing the sampled data which is then transmitted to a PC using UART communication protocol for further processing and analysis.

**Minimum Detectable Amount of Nanomag50 and SHB30 MNPs by One-stage Lock-in MPS system.** The sensitivity of our one-stage lock-in MPS system was firstly evaluated by assessing the minimum amount of MNPs detectable in liquid phase. Briefly, two-fold dilutions of Nanomag50 and SHB30 MNP samples are prepared in glass vials and each vial contains 80 μL MNP suspension. Details of the experimental designs can be found in Supporting Information S3.

The Nanomag50 MNPs were diluted up to 8192 times, from 5 mg/mL (400 μg per vial, no dilution) to 610 ng/mL (48.8 ng per vial, 8192-fold dilution). Three independent MPS readings were carried out on each sample. As shown in Figure 3, the amplitudes of higher harmonics (i.e., from the 3$^{rd}$ to the 15$^{th}$ harmonics) linearly decreased as the amount of Nanomag50 MNPs decreases. The inset of Figure 3(g) shows a zoomed in view of the 3$^{rd}$ harmonic amplitude collected from lowest amount of Nanomag50 MNPs. It was concluded that the minimum detectable amount of Nanomag50 MNPs by MPS system is 781 ng (512-fold dilution).

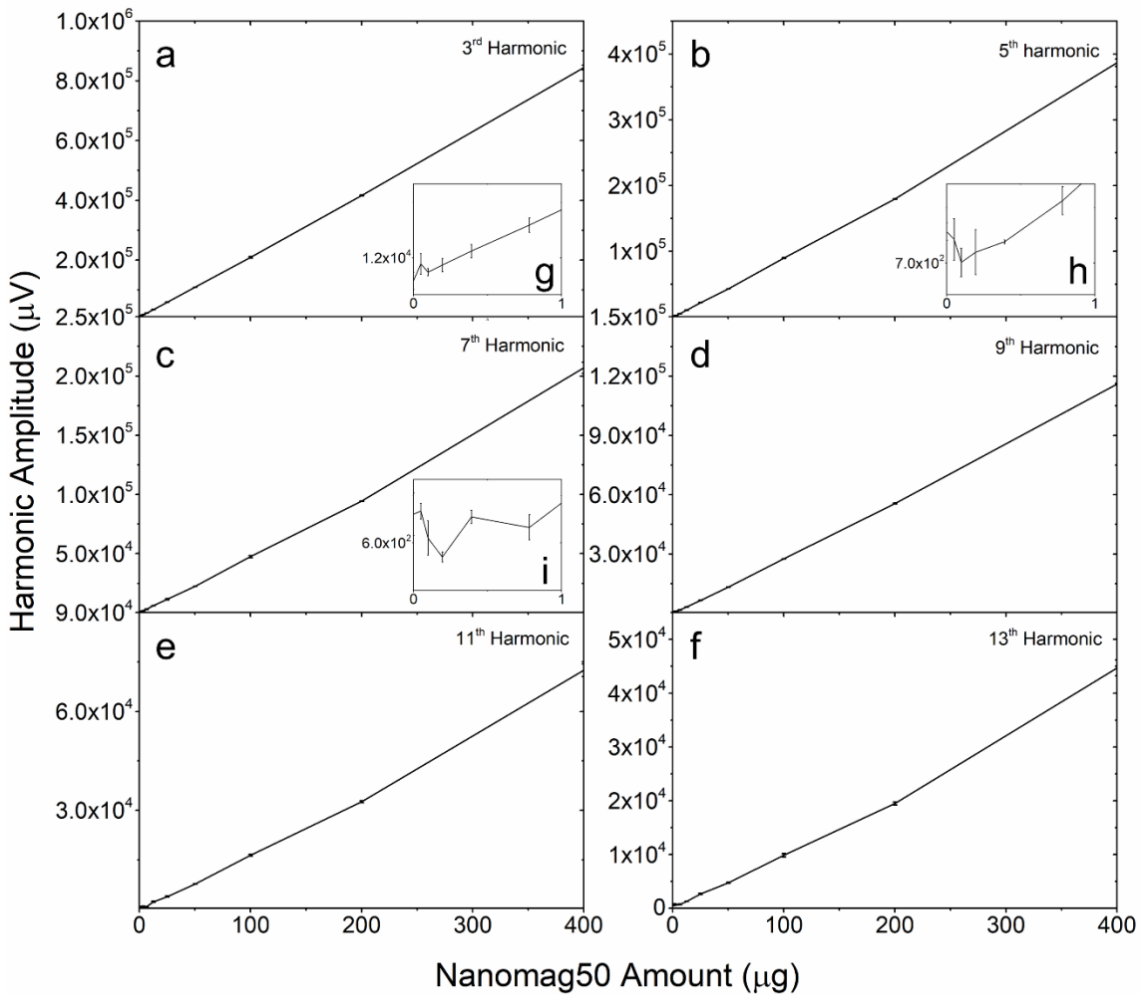

Figure 3. The (a) 3$^{rd}$, (b) 5$^{th}$, (c) 7$^{th}$, (d) 9$^{th}$, (e) 11$^{th}$, and (f) 15$^{th}$ harmonics collected from varying amount of Nanomag50 MNPs. (g) – (i) are the zoom in view of the 3$^{rd}$, 5$^{th}$, and 7$^{th}$ harmonics collected from samples with MNP amount below 1 µg per vial. Error bars represent standard errors.

In addition, varying amount of SHB30 MNPs were prepared by two-fold dilutions in the same manner. The SHB30 MNPs are diluted up to 2048 times, from 1 mg/mL (80 µg per vial, no dilution) to 488 ng/mL (39 ng per vial, 2048-fold dilution). As shown in Figure 4, the amplitude of higher harmonic linearly decreases as the SHB30 amount decreases in the samples. The minimum detectable amount of SHB30 MNPs by the developed one-stage lock-in topology based MPS system is 78 ng (1024-fold dilution). The modified MPS system shows a 50-fold improvement in sensitivity for detection of iron oxide MNPs when compared to our previous work (limit of detection was 4 µg) not utilizing lock-in based approach as shown in Figure 1(d).[19]

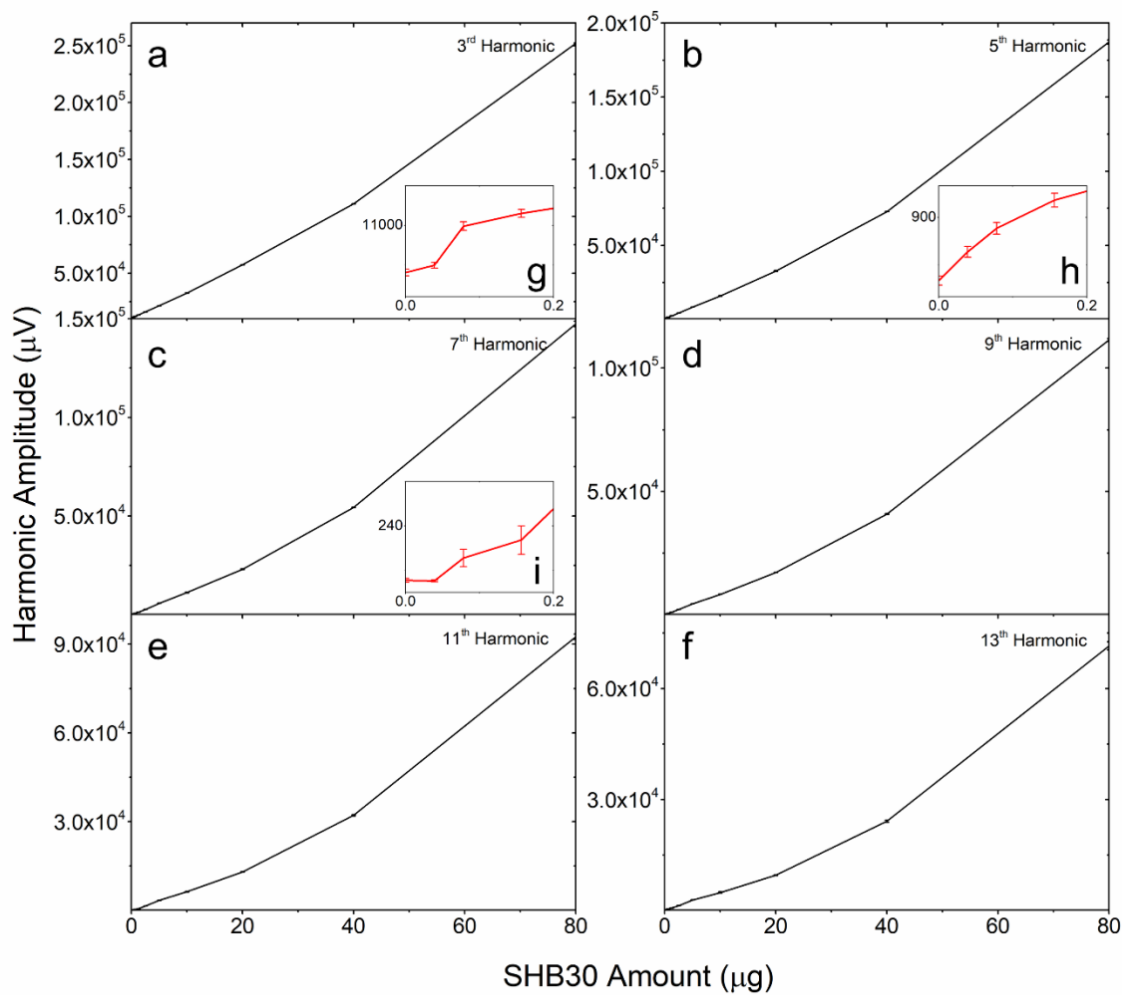

Figure 4. The (a) 3$^{rd}$, (b) 5$^{th}$, (c) 7$^{th}$, (d) 9$^{th}$, (e) 11$^{th}$, and (f) 15$^{th}$ harmonics collected from varying amount of SHB30 MNPs. (g) – (i) are the enlarged view of the 3$^{rd}$, 5$^{th}$, and 7$^{th}$ harmonics collected from samples with MNP amount below 200 ng per vial. Error bars represent standard error.

**Effect of MNP Amount on Analyte Concentration–Response Profiles.** Herein, we explored the effect of MNP amount on concentration-response curves for analyte detection by using a streptavidin-biotin binding system. As shown in Scheme 1, SHB30 MNPs were surface functionalized with biotin molecules that have high binding affinity to streptavidin. When there is an inadequate amount (or lack of) of streptavidin to bind MNPs, as shown in Scheme 1, MNPs can freely rotate through Brownian relaxation (Scheme 1(b)) to align with driving magnetic fields, thus, showing strong dynamic magnetic responses and large harmonic amplitudes (Scheme 1(c)&(d)). With increasing amount of streptavidin in the sample, MNPs form cluster matrices and exhibit larger hydrodynamic sizes (the closed loops in Scheme 1). As a result, the clustered MNPs lose their rotational freedom and exhibit weaker dynamic magnetic respons to the driving fields and consequently lower harmonic amplitudes are expected (see Scheme 1(d)). Based on the relative abundance of MNPs and streptavidin in the liquid sample, the degree of MNP clustering goes through zone of inadequate analyte (a.1), linear response region (a.2), and zone of excessive analyte (a.3, a.4) as labeled in Scheme 1. In the linear response region, the amounts of MNPs and streptavidin reach concentrations that enable formation of analyte-MNP cluster matrices. The formation of the matrix increases in a linear ratio until it exceeds an optimal concentration when the matrices dissociate, with excessive amount of streptavidin. This MNP inter-linking continues along a linear scale until formation of large clusters of MNP-streptavidin completely blocks the Brownian relaxation of MNPs, resulting in the lowest harmonic amplitudes. As the concentration of streptavidin (analyte) increases, MNP-biotin binding sites will become saturated until they are fully occupied by streptavidin (see Scheme 1 (a.4)), the large cluster matrices are disrupted, and MNP clusters no longer form the majority in liquid phase. Individual MNPs when saturated with streptavidin in this scenario will exhibit harmonic response that is greater than that of large clusters but smaller than that of the free MNPs due to an increment in effective hydrodynamic size.

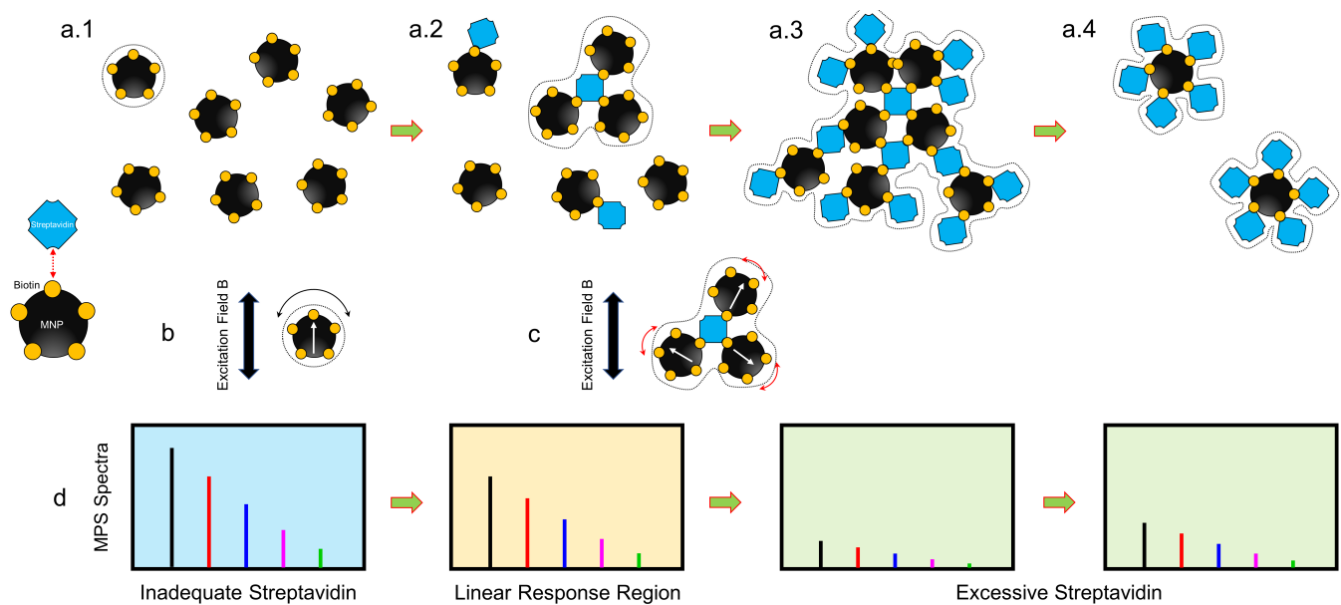

Scheme 1. Streptavidin has a high binding affinity to biotin on the surface of SHB30 MNPs. With the addition of streptavidin as shown in (a), the MNPs form clusters and their Brownian relaxation process is blocked, (b) & (c). Under different scenarios with inadequate streptavidin, linear response region, and excess streptavidin, the dynamic magnetic responses in the form of MPS spectra (higher harmonics from MNPs) become weaker, (d). The dashed closed loops depict the hydrodynamic sizes.

The goal of this experiment was to examine the effect of varying MNP amounts in the assay on the concentration-response profiles for detection of streptavidin in liquid phase. Four experimental groups were designed, each consisting of 10 samples/vials and each vial containing 80 µL of SHB30 MNPs with concentration: 1) 8.5 nM (group I, particle amount: 680 fmole per vial, 4-fold dilution), 2) 4.25 nM (group II, particle amount: 340 fmole per vial, 8-fold dilution), 3) 2.125 nM (group III, particle amount: 170 fmole per vial, 16-fold dilution), and 4) 1.0625 nM (group IV, particle amount: 85 fmole per vial, 32-fold dilution). To all 10 vials from each group, 80 µL of varied streptavidin concentrations ranging from 400 nM to 0 nM was added, as shown in Table 1.

Table 1. Experimental Designs for Groups I – IV.

| Group No. Sample Index | SHB30 MNP Concentration/Amount (80 μL per vial) | Streptavidin Concentration/Amount (80 μL per vial) |
| --- | --- | --- |
| Group I #1-10 | 8.5 nM (4-fold dilution), 680 fmole | 400 nM, 32 pmole (#1) |
|  |  | 300 nM, 24 pmole (#2) |
| Group II #1-10 | 4.25 nM (8-fold dilution), 340 fmole | 200 nM, 16 pmole (#3) |
|  |  | 150 nM, 12 pmole (#4) |
| Group III #1-10 | 2.125 nM (16-fold dilution), 170 fmole | 100 nM, 8 pmole (#5) |
|  |  | 50 nM, 4 pmole (#6) |
| Group IV #1-10 | 1.0625 nM (32-fold dilution), 85 fmole | 25 nM, 2 pmole (#7) |
|  |  | 10 nM, 800 fmole (#8) |
|  |  | 5 nM, 400 fmole (#9) |
|  |  | 0 nM, 0 fmole (#10) |

In short, 80 μL SHB30 MNP suspensions of varying dilutions were mixed with 80 μL streptavidin of varying concentrations. The mixtures are incubated at room temperature for 30 min on a shaker, to allow the binding of biotins on SHB30 MNPs to streptavidin in the liquid phase. MPS measurements are carried out on each vial and six independent MPS readings are taken from each sample.

As shown in Figure 5(a), clear concentration-response curves are observed from all four groups, the $3^{rd}$ harmonics gradually decrease as the amount of streptavidin increases from vial #10 to #1. Vials #1 to #4 for group IV demonstrates the phenomena of MNP-biotin saturation with excess streptavidin, where most MNP binding sites are fully occupied with streptavidin decreasing the formation of large MNP clusters. This phenomenon is most apparent when higher $3^{rd}$ and $5^{th}$ harmonic signals were observed at higher analyte concentrations compared to vials #5-7 with lower streptavidin. It should also be noted that $3^{rd}$ harmonic amplitude in vials #1- #4 (streptavidin saturation) was smaller than that of vials #9-10 (inadequate streptavidin case) where unbound MNPs form majority in the liquid phase. For experimental groups I-III, the higher amount of streptavidin in vials #1-5 results in the clustering of SHB30 MNPs and hence lower harmonic signals are observed. As a result, this clustering hinders the Brownian relaxation of MNPs to realign their magnetic moments to the driving fields. As we reduce the streptavidin concentration/amount, it reaches to linear response region where the relative number of biotins from SHB30 MNP is in the same order of magnitude of the streptavidin. With further reduction of streptavidin concentration/amount, the harmonic signal increases and reaches a state where SHB30 MNPs can freely rotate following the driving magnetic fields since there is inadequate amount of streptavidin to enable MNP clustering and hinder their Brownian relaxation. We observed a weak reversal of harmonic signals from its nadir for groups I to III (in vials #1-4), however the phenomenon was not as prominent as observed for group IV, even at the highest (400 nM) concentration of streptavidin tested. It is plausible that 400 nM streptavidin is insufficient to completely saturate MNP associated biotins in groups I – III. These three different response zones are highlighted in blue (excessive amount of streptavidin), orange (linear response region), and green (inadequate streptavidin) regions in Figure 5(e-h) as well as in Scheme 1(d). Overall, the higher harmonics show parallel concentration-response curves in the $3^{rd}$ harmonic signals. It is observed that the linear response region of concentration-response curve moves towards lower streptavidin quantities (as shown by the arrow in Figure 5(a) and summarized in Table 2) with the dilution of SHB30 MNPs from group I to group IV. Figure 5(b-d) compares the $3^{rd}$ harmonic amplitudes of SHB30 MNPs from groups I – IV detecting the same concentration/amount of streptavidin.

Table 2. Linear Response Region Observed from Concentration-response Profiles Based on Different MNP Concentrations

| Group Index | SHB30 MNP Concentration | Linear Response Region (Streptavidin Concentration) |
|---|---|---|
| I   | 8.5 nM    | 50 – 100 nM |
| II  | 4.25 nM   | 25 – 100 nM |
| III | 2.125 nM  | 10 – 50 nM  |
| IV  | 1.0625 nM | 5 – 25 nM   |

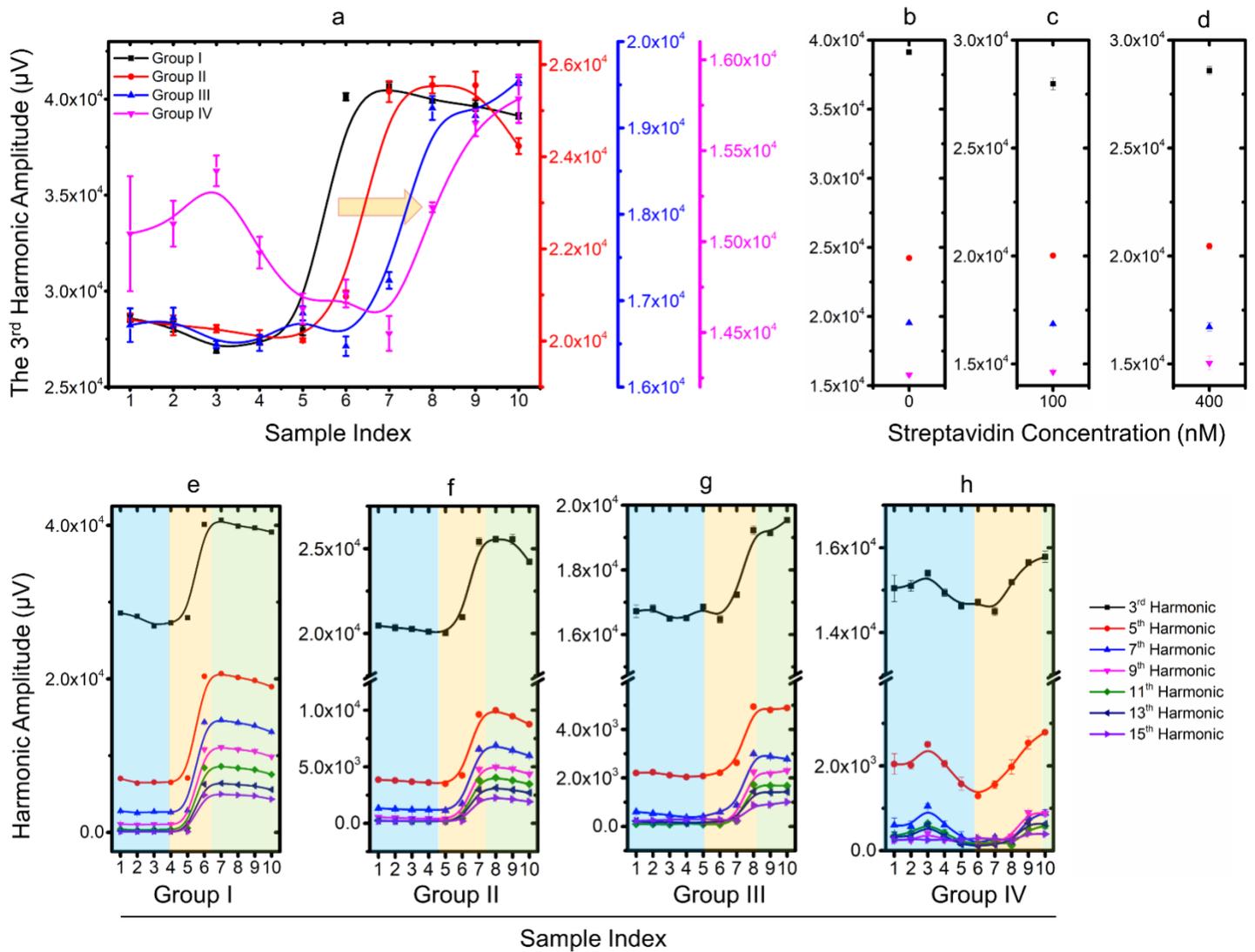

Figure 5. (a) The concentration-response profiles for streptavidin based on the 3rd harmonic amplitudes of SHB30 MNPs with 8.5 nM (group I, black), 4.25 nM (group II, red), 2.125 nM (group III, blue), and 1.0625 nM (group IV, magenta) concentrations. (b) – (d) are the 3rd harmonics of different dilutions of SHB30 MNPs mixed with 0 nM, 100 nM, and 400 nM streptavidin, respectively. (e) – (h) are the concentration-response profiles for streptavidin from groups I – IV, based on the 3rd to the 15th harmonics. Error bars represent standard errors.

Interestingly, the best SNR performance was not observed from the 3rd harmonic response of the MNPs but, instead the least standard deviation was observed consistently when 9th and 15th harmonic were used to represent MNP binding information. Figures 6(a) & (b) present the concentration-response profiles for streptavidin based on the 9th and 15th harmonics. A plausible explanation for the observation could be attributed to higher 1/f noise for the 3rd harmonic signal. During experiments it was observed that 3rd harmonic was more susceptible to surrounding noises. Most of these issues can be addressed by providing proper magnetic and EMI isolation to the system along with implementation of a second lock-in corresponding to the harmonic of interest to reject phase separable noise.

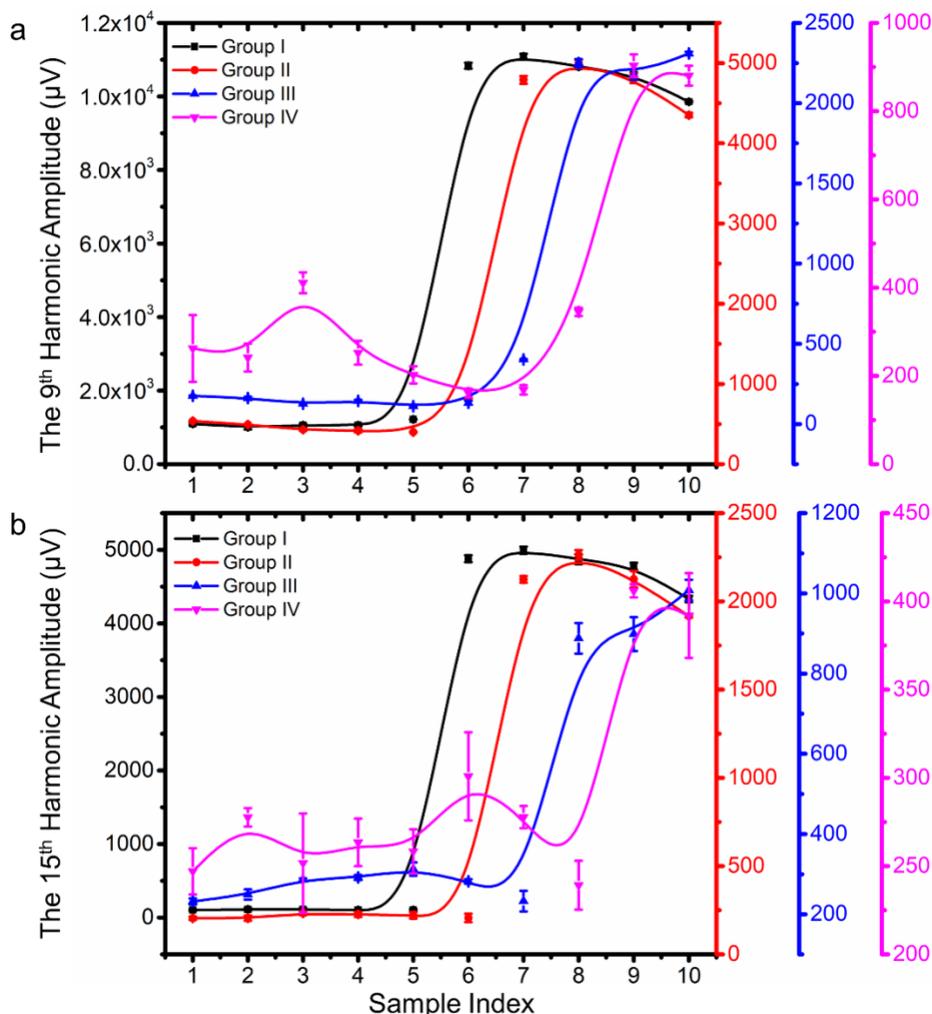

Figure 6. The concentration-response profiles for streptavidin based on (a) the 9$^{th}$ and (b) the 15$^{th}$ harmonic amplitudes of SHB30 MNPs with 8.5 nM (group I, black), 4.25 nM (group II, red), 2.125 nM (group III, blue), and 1.0625 nM (group IV, magenta) concentrations. Error bars represent standard errors.

The streptavidin-biotin binding experiments were also carried out on multi-core Nanomag50 MNPs, where four experimental groups each consists of 10 samples/vials were designed and each vial contained 80 μL Nanomag50 MNP of 8.5 nM, 4.25 nM, 2.125 nM, and 1.0625 nM concentrations. Due to the characteristic of the multi-core Nanomag50 MNPs having smaller superparamagnetic iron oxide nanoparticles embedded in the matrix, their Brownian relaxation is intrinsically blocked, and hence, clustering dose not impact their dynamic magnetic response. Thus, very weak or no harmonic signal changes were observed from Nanomag50 MNP samples mixed with varying concentration/amount of streptavidin. The experimental design and results of multi-core Nanomag50 MNP-based streptavidin tests are summarized in Supporting Information S4. The results indicate that multi-core MNPs, whose Brownian relaxation is intrinsically blocked cannot be used for liquid phase MPS-based bioassays.

## CONCLUSIONS AND FUTURE PROSPECTS

In the present study, we have reported a method for passive cancellation of feedthrough signal for dual-frequency (2 AC driving fields) MPS methodology. The sensitivity of this one-stage lock-in MPS system is first evaluated by detecting two-fold dilutions of commercial iron oxide MNPs: SHB30 and Nanomag50. The lowest amount detectable by the system was confirmed to be 78 ng for the single-core SHB30 and 781 ng for the multi-core Nanomag50 iron oxide MNPs. In addition, using a streptavidin-biotin binding system as a model, we explored the effects of MNP amount on concentration-responses profiles for detecting target analytes. By fine tuning the MNP amount/concentration in the sample, we were able to shift the linear response region for streptavidin detection. Results confirmed that the liquid phase bioassay scheme shows improved sensitivity on our one-stage lock-in MPS system when lower amount/concentration of MNPs is used. The linear response region shifts detection ability to towards lower concentration of streptavidin using lower MNP quantities. Using 8.5 nM, 4.25 nM, 2.125 nM, and 1.0625 nM concentrations of MNPs for streptavidin detection, the linear response region shifts from 50 – 100 nM, 25 – 100 nM, 10 – 50 nM, and down to 10 – 25 nM respectively, pushing the detection limit of streptavidin further into the femtomole range. These concentration-response profiles indicate the possibility for improved bioassay sensitivities by using lower amount/concentration of MNPs. In the present study, our one-stage lock-in MPS system is able to detect as low as 800 fmole of streptavidin using 1.0625 nM concentration of MNPs. However, this detection limit can be further improved by using lower amounts (higher dilutions) of MNPs. In addition to improved sensitivities, the cost per assay can be further reduced by using less amount of MNPs. Low-cost options allow the point-of-care assays to be available in impoverished regions with scarce medical resources.

Our future plans include improving the sensitivity of MPS system by using active feedthrough cancellation techniques. The passive feedthrough cancellation reported in the present study help remove the feedthrough signals before ADC sampling. Therefore, with proper amplifications in place, we can take advantage of true ADC resolution for improved sensitivity. However, this method of removing feedthrough signals and amplifying remnant signals still holds disadvantage as the intrinsic noise components also get amplified through the amplification units. Active cancellation of feedthrough can help improve the signal-to-noise ratio before the instrumentation amplifier stage and hence allow for better sensitivity in MPS bioassay applications. We believe this passive feedthrough cancellation methodology in MPS system design and fine tuning MNP amount/concentration to shift linear response region will elucidate new ways of increasing detection sensitivity of MPS-based bioassays.

▪ **ASSOCIATED CONTENT**

**Supporting Information:** S1. Magnetic Properties of Nanomag50 and SHB30 MNPs.; S2. Low noise powerline split implementation.; S3. Experimental Designs for Varying Amount of Nanomag50 and SHB30 MNPs.; S4. Multi-core Nanomag50 MNPs for Streptavidin Detection.

- **AUTHOR INFORMATION**

**Corresponding Authors**

*E-mail: wuxx0803@umn.edu (K. W.)

*E-mail: cheeran@umn.edu (M. C-J. C)

*E-mail: jpwang@umn.edu (J.-P. W.)

**ORCID**

Vinit Kumar Chugh: 0000-0001-7818-7811

Kai Wu: 0000-0002-9444-6112

Venkatramana D. Krishna: 0000-0002-1980-5525

Arturo di Girolamo: 0000-0002-6906-8754

Robert P. Bloom: 0000-0002-7781-5270

Yongqiang Andrew Wang: 0000-0003-2132-2490

Renata Saha: 0000-0002-0389-0083

Shuang Liang: 0000-0003-1491-2839

Maxim C-J Cheeran: 0000-0002-5331-4746

Jian-Ping Wang: 0000-0003-2815-6624

**Author Contributions**

[†]V.K.C. and K.W. have contributed equally to this work.

**Notes**

The authors declare no conflict of interest.

- **ACKNOWLEDGMENTS**

This study was financially supported by the Institute of Engineering in Medicine, the Robert F. Hartmann Endowed Chair professorship, the University of Minnesota Medical School, and the University of Minnesota Physicians and Fairview Health Services through COVID-19 Rapid Response Grant. This study was also financially supported by the U.S. Department of Agriculture - National Institute of Food and Agriculture (NIFA) under Award Number 2020-67021-31956. Research reported in this publication was supported by the National Institute of Dental & Craniofacial Research of the National Institutes of Health under Award Number R42DE030832. The content is solely the responsibility of the authors and does not necessarily represent the official views of the National Institutes of Health. Portions of this work were conducted in the Minnesota Nano Center, which is supported by the National Science Foundation through the National Nano Coordinated Infrastructure Network (NNCI) under Award Number ECCS-1542202.

TOC:

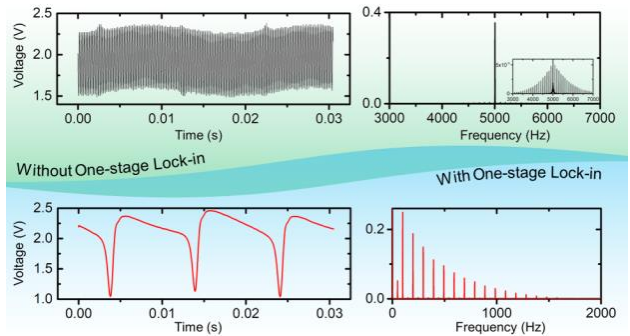

# Supporting Information

**Magnetic Particle Spectroscopy (MPS) with One-stage Lock-in Implementation for Magnetic Bioassays with Improved Sensitivities**

Vinit Kumar Chugh[a,†], Kai Wu[a,†,*], Venkatramana D. Krishna[b], Arturo di Girolamo[a], Robert P. Bloom[a], Yongqiang Andrew Wang[c], Renata Saha[a], Shuang Liang[d], Maxim C-J Cheeran[b,*], and Jian-Ping Wang[a,*]

[a]Department of Electrical and Computer Engineering, University of Minnesota, Minneapolis, MN 55455, United States

[b]Department of Veterinary Population Medicine, University of Minnesota, St. Paul, MN 55108, United States

[c]Ocean Nano Tech LLC, San Diego, CA 92126, United States

[d]Department of Chemical Engineering and Material Science, University of Minnesota, Minneapolis, MN 55455, United States

[†]V.K.C. and K.W. have contributed equally to this work.

*E-mails: wuxx0803@umn.edu (K. W.); cheeran@umn.edu (M. C-J. C); jpwang@umn.edu (J.-P. W.)

## S1. Magnetic Properties of Nanomag50 and SHB30 MNPs.

The static magnetic hysteresis loops of Nanomag50 and SHB30 MNPs are measured by PPMS and shown in Figure S1. External magnetic fields are swept from -5000 Oe to +5000 Oe and -500 Oe to +500 Oe. The saturation magnetizations ($M_s$, emu/g) of Nanomag50 and SHB30 MNPs under 5000 Oe field are 30.7 emu/g and 36.8 emu/g, respectively. The specific magnetic magnetizations (M, emu/g) of Nanomag50 and SHB30 MNPs under 500 Oe field are 18.2 emu/g and 29.2 emu/g, respectively. Nanomag50 MNPs show superparamagnetic properties with zero magnetic coercivity while, on the other hand, SHB30 shows a coercivity field of 36 Oe. Due to the increased inter-particle distances introduced by surface functional groups (biotin in this case for SHB30), this negligible magnetic coercivity won't cause SHB30 to cluster in the absence of magnetic fields. In addition, the magnetic moment per particle (m, emu/particle) is also calculated based on the particle concentrations, as shown in Table S1.

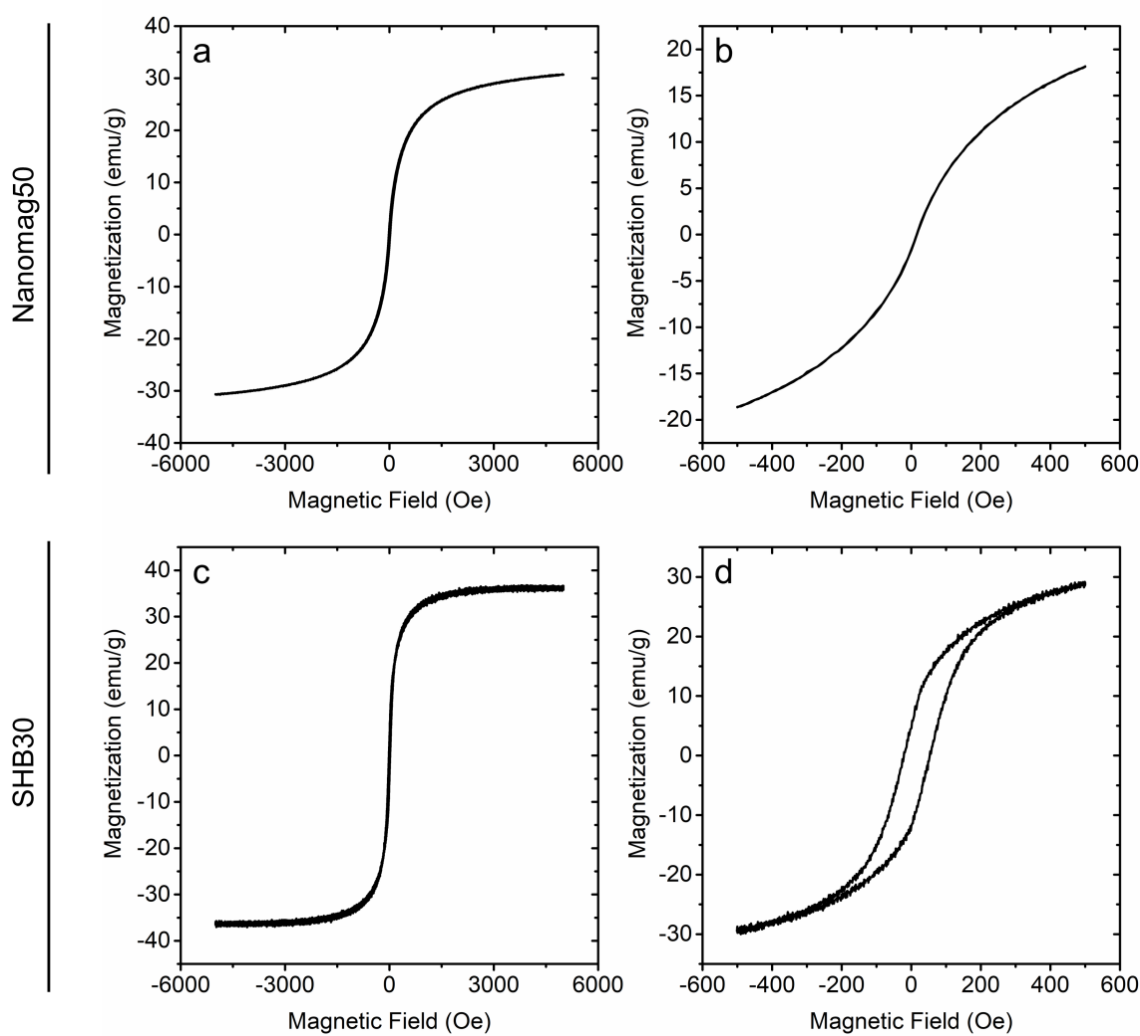

Figure S1. The static magnetic hysteresis loops of Nanomag50 and SHB30 nanoparticles measured by PPMS with external magnetic field ranges of (a) & (c) 5000 Oe and (b) & (d) 500 Oe, respectively.

Table S1. Magnetic properties of Nanomag50 and SHB30 nanoparticles.

| MNP | Concentration | Ms (5000 Oe) emu/g | m (5000 Oe) emu/particle | M (500 Oe) emu/g | m (500 Oe) emu/particle | Coercivity Oe |
|---|---|---|---|---|---|---|
| Nanomag50 | 5 mg/mL (weight) 91.4 nM (particle) | 30.7 | $2.79 \times 10^{-15}$ | 18.2 | $1.65 \times 10^{-15}$ | 0 |
| SHB30 | 1 mg/mL (weight) 34 nM (particle) | 36.8 | $1.75 \times 10^{-15}$ | 29.2 | $1.41 \times 10^{-15}$ | 36 |

## S2. Low noise powerline split implementation.

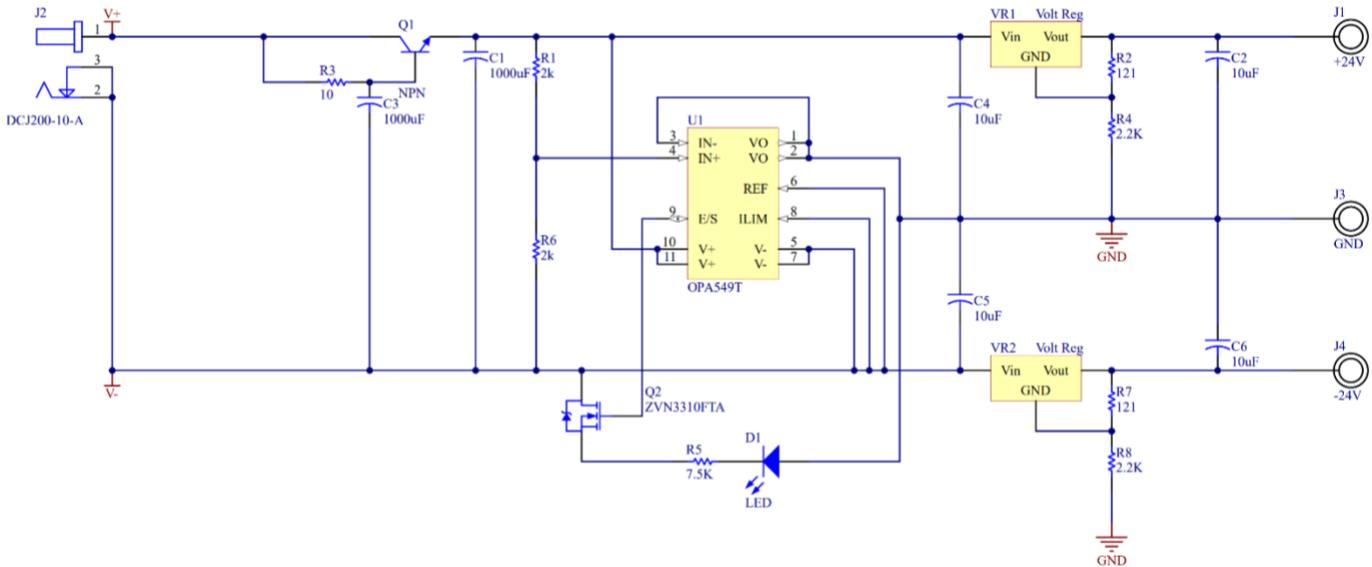

Figure S2. Schematic for low-noise voltage generation of ±24 V for the excitation coil driver setup.

## S3. Experimental Designs for Varying Amount of Nanomag50 and SHB30 MNPs.

A total of 15 vials containing two-fold dilutions of Nanomag50 multi-core MNPs are prepared, with concentrations from 5 mg/mL (vial #1) down to 610 ng/mL (vial #14). Vial #15 is a negative control group without loading any MNPs. Each vial contains 80 μL of MNP solution. The Nanomag50 MNP amount per vial as well as dilution folds are summarized in Table S2.

Table S2. Experimental Designs for Two-fold Dilutions of Nanomag50 MNPs.

| Vial # | Nanomag50 Concentration (mg/mL) | MNP Amount (ug) | Dilution |
|---|---|---|---|
| 1 | 5 | 400 | 1 |
| 2 | 2.5 | 200 | 2 |
| 3 | 1.25 | 100 | 4 |
| 4 | 0.625 | 50 | 8 |
| 5 | 0.3125 | 25 | 16 |
| 6 | 0.15625 | 12.5 | 32 |
| 7 | 0.078125 | 6.25 | 64 |
| 8 | 0.0390625 | 3.125 | 128 |
| 9 | 0.01953125 | 1.5625 | 256 |
| 10 | 0.009765625 | 0.78125 | 512 |
| 11 | 0.004882813 | 0.390625 | 1024 |
| 12 | 0.002441406 | 0.1953125 | 2048 |
| 13 | 0.001220703 | 0.09765625 | 4096 |
| 14 | 0.000610352 | 0.048828125 | 8192 |
| 15 | 0 | 0 | - |

A total of 13 vials containing two-fold dilutions of SHB30 single-core MNPs are prepared, with concentrations from 1 mg/mL (vial #1) down to 488 ng/mL (vial #12). Vial #13 is a negative control group without loading any MNPs. Each vial contains 80 μL of MNP solution. The SHB30 MNP amount per vial as well as dilution folds are summarized in Table S3.

Table S3. Experimental Designs for Two-fold Dilutions of SHB30 MNPs.

| Vial # | SHB30 Concentration (mg/mL) | MNP Amount (ug) | Dilution |
|---|---|---|---|
| 1 | 1 | 80 | 1 |
| 2 | 0.5 | 40 | 2 |
| 3 | 0.25 | 20 | 4 |
| 4 | 0.125 | 10 | 8 |
| 5 | 0.0625 | 5 | 16 |
| 6 | 0.03125 | 2.5 | 32 |
| 7 | 0.015625 | 1.25 | 64 |

| 8 | 0.0078125 | 0.625 | 128 |
| 9 | 0.00390625 | 0.3125 | 256 |
| 10 | 0.001953125 | 0.15625 | 512 |
| 11 | 0.000976563 | 0.078125 | 1024 |
| 12 | 0.000488281 | 0.0390625 | 2048 |
| 13 | 0 | 0 | - |

**S4. Multi-core Nanomag50 MNPs for Streptavidin Detection.**

Four experimental groups each consists of 10 samples/vials are designed and each vial contains 80 μL Nanomag50 MNP of 10.75-, 21.5-, 43-, and 86-fold dilutions are designed as shown in Table S4. The concentration-response profiles for streptavidin from groups A – D are plotted in Figure S3.

Table S4. Experimental Designs for Groups A – D.

| Group Index Sample Index | Nanomag50 MNP Concentration/Amount (80 μL per vial) | Streptavidin Concentration/Amount (80 μL per vial) |
|---|---|---|
| Group A #1-10 | 8.5 nM (10.75-fold dilution), 680 fmole | 400 nM, 32 pmole (#1) |
| | | 200 nM, 16 pmole (#2) |
| Group B #1-10 | 4.25 nM (21.5-fold dilution), 340 fmole | 100 nM, 8 pmole (#3) |
| | | 50 nM, 4 pmole (#4) |
| Group C #1-10 | 2.125 nM (43-fold dilution), 170 fmole | 25 nM, 2 pmole (#5) |
| | | 12.5 nM, 1 pmole (#6) |
| Group D #1-10 | 1.0625 nM (86-fold dilution), 85 fmole | 6.25 nM, 500 fmole (#7) |
| | | 3.13 nM, 250 fmole (#8) |
| | | 1.56 nM, 125 fmole (#9) |
| | | 0 nM, 0 fmole (#10) |

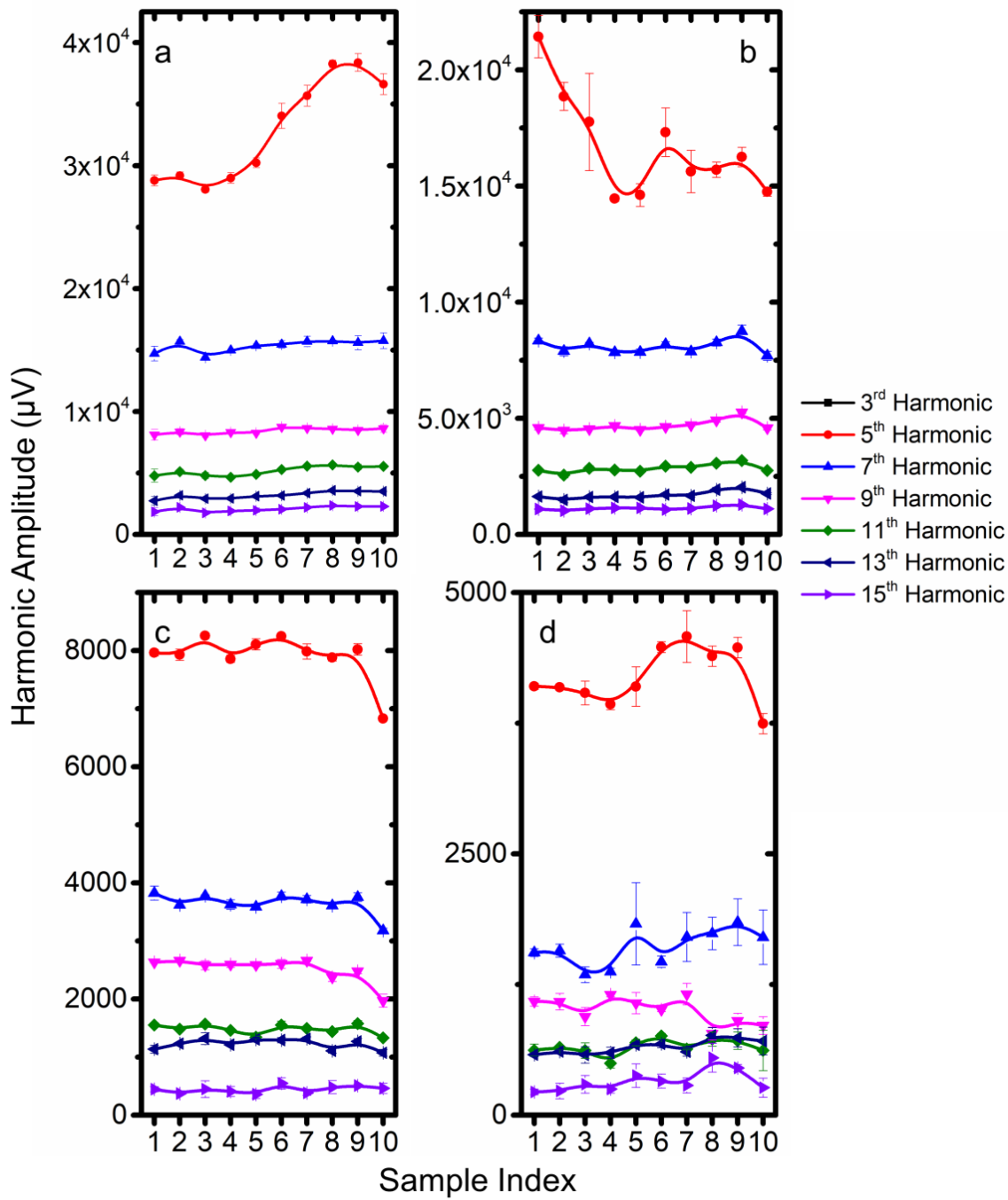

Figure S3. Concentration-response profiles for streptavidin from groups A - D, based on the 3rd to the 15th harmonics. Error bars represent standard errors.